\documentclass[conference]{IEEEtran}
\usepackage{cite}
\usepackage{tabularx}
\IEEEoverridecommandlockouts
\usepackage{optidef}
\RenewEnviron{BaseMiniExclam}[7]{
  \selectConstraintMult{#1}
  \renewcommand{\localOptimalVariable}{#2}
  \begin{subequations}
  \ifthenelse{\equal{#7}{b}}{\allowdisplaybreaks}
  \begin{alignat}{5}
    \bodyobj{#2}{#3}{#6}{#5}
    \BODY
  \end{alignat}
  \end{subequations}%
  \setStandardMini
}
\usepackage{balance}
\usepackage[algo2e]{algorithm2e} 
\SetKwInput{KwInput}{Input}              
\SetKwInput{KwOutput}{Output}             
\ifCLASSINFOpdf
  \usepackage[pdftex]{graphicx}
  \graphicspath{{../pdf/}{../jpeg/}}
  \DeclareGraphicsExtensions{.pdf,.jpeg,.png}
\else
  \usepackage[dvips]{graphicx}
  \graphicspath{{../eps/}}
  \DeclareGraphicsExtensions{.eps}
\fi
\usepackage{amssymb} 
\usepackage{epsfig}
\usepackage{svg}
\usepackage{amsmath,tikz}
\usepackage{amsfonts}
\usepackage[T1]{fontenc}
\usepackage{eqparbox}
\usepackage{graphicx}
\usepackage{array}
\usepackage{epstopdf}
\usepackage{amsthm}
\usepackage{comment}
\usepackage{algorithm}
\usepackage{color,soul}
\usepackage{enumitem}
\usepackage{csquotes}
\usepackage{soul}
\usepackage{tabularx}
\usepackage{multicol}
\usepackage{longtable}
\usepackage{lipsum}
\usepackage{bigfoot}
\usepackage[noend]{algpseudocode}
\usepackage{rotating}
\usepackage[hidelinks]{hyperref}
\usepackage{color,array}
\usepackage{bm}
\usepackage{amssymb}
\usepackage{lettrine}
\usepackage{xcolor, soul}
\usepackage{cite}
\usepackage{amsmath}
\usepackage{array}
\newcolumntype{L}[1]{>{\raggedright\let\newline\\\arraybackslash\hspace{0pt}}m{#1}}
\newcolumntype{C}[1]{>{\centering\let\newline\\\arraybackslash\hspace{0pt}}m{#1}}
\newcolumntype{R}[1]{>{\raggedleft\let\newline\\\arraybackslash\hspace{0pt}}m{#1}}
\usepackage{url}
\usepackage{xcolor}
\usepackage[normalem]{ulem}

\usepackage[acronym,nonumberlist,style=super,toc]{glossaries}

\newacronym{3gpp}{3GPP}{3rd generation partnership project}

\newacronym{awgn}{AWGN}{Additive White Gaussain Noise}
\newacronym{af}{AF}{artificial fading}

\newacronym{bsc}{BSC}{binary symmetric channel}
\newacronym{bs}{BS}{base station}
\newacronym{cr}{CR}{channel reciprocity}
\newacronym{crc}{CRC}{cyclic redundancy check}
\newacronym{ci}{CI}{cooperation information}

\newacronym{doss}{DOSS}{difference of signal strength}
\newacronym{dqn}{DQN}{deep Q-network}
\newacronym{drl}{DRL}{deep reinforcement learning}
 \newacronym{ddpg}{DDPG}{deep deterministic policy gradient}
\newacronym{gkg}{GKG}{group key generation}
\newacronym{gkd}{GKD}{group key disagreement}
\newacronym{gk}{GK}{group key }
\newacronym{gsk}{GSK}{group secret key}
\newacronym{los}{LoS}{line-of-sight}

\newacronym{nc}{NC}{network coding}
\newacronym{nlos}{NLoS}{non line-of-sight}

\newacronym{puf}{PUF}{physically unclonable function}
\newacronym{pls}{PLS}{physical layer security}
\newacronym{pkg}{PKG}{pairwise key generation}
\newacronym{pufe}{PUFe}{\gls{puf} emulator}

\newacronym{qos}{QoS}{quality-of-service}
\newacronym{rssi}{RSSI}{received signal strength indicator}
\newacronym{rl}{RL}{reinforcement learning}
\newacronym{ss}{SS}{secure sketch}
\newacronym{sgkg}{SGKG}{sequential group key generation}
\newacronym{ssgk}{SSGK}{sequential secret group key}

\newacronym{sinr}{SINR}{signal-to-interference plus noise}
\newacronym{sgkgr}{SGKGR}{sequential group key generation with redundancy}

\newacronym{uav}{UAV}{unmanned aerial vehicle}
\newacronym{ber}{BER}{bit error rate}
\newacronym{csi}{CSI}{channel state informatiom}

\newacronym{tdd}{TDD}{time-division duplexing}

\newacronym{tdma}{TDMA}{time division multiple access}
\newacronym{id}{ID}{identity number}

\newacronym{xor}{XOR}{exclusive-OR}
\newacronym{arq}{ARQ}{automatic repeat request}

\newacronym{rss}{RSS}{received signal strength}

\newacronym{a2a}{A2A}{air-to-air}

\newacronym{slsa}{SLSA}{successful link selection algorithm}

\title{Entanglement Request Scheduling in Quantum Networks Using Deep Q-Network}

\author{\IEEEauthorblockN{Gongyu Ni\IEEEauthorrefmark{1}, %\IEEEauthorrefmark{2},
Lester Ho\IEEEauthorrefmark{1}, 
Holger Claussen\IEEEauthorrefmark{1}\IEEEauthorrefmark{2}\IEEEauthorrefmark{3}}
\IEEEauthorblockA{\IEEEauthorrefmark{1}Tyndall National Institute, Dublin, Ireland}
\IEEEauthorblockA{\IEEEauthorrefmark{2}University College Cork, Ireland}
\IEEEauthorblockA{\IEEEauthorrefmark{3}Trinity College Dublin, Ireland}
Emails: 
\{gongyu.ni, lester.ho, holger.claussen\}@tyndall.ie\vspace{-2em} 
}

\begin{document}

\maketitle
\begin{abstract}
In this paper, a novel Deep Q-Network (\text{DQN}) based scheduling method to optimize delay time and fairness among entanglement requests in quantum repeater networks is proposed. %These entanglement requests aim to establish entanglement between two end nodes. 
The scheduling of requests determines which pairs of end nodes should be entangled during the current time slot, while other pairs are placed in a queue for future slots. 
% However, existing research on quantum networking often relies on simple statistical models to capture the behavior of quantum hardware, such as the failure rate of establishing entanglement. 
% Moreover, current quantum simulators do not support network behaviors, including handling, pending, and dropping requests. 
% To bridge the gap between quantum deployments and network behaviors, in this paper a dynamic network model is presented, encompassing quantum simulations, random topologies, and user modeling. 
However, existing research on quantum networking often relies on simple statistical models to capture the behavior of quantum hardware, such as the failure rate of establishing entanglement. 
Moreover, current quantum simulators do not support network behaviors, including handling, pending, and dropping requests. 
To bridge the gap between quantum deployments and network behaviors, in this paper a dynamic network model is presented, encompassing quantum simulations, random topologies, and user modeling. 
The DQN based scheduling scheme allows us to balance the conflicting objectives of minimizing delay time and maximizing fairness among these entanglement requests. The proposed technique was evaluated using simulations, with results showing that the proposed \text{DQN} achieves higher performance compared to Greedy, Proportional fair and FIFO scheduling schemes.
\end{abstract}
\glsresetall

\glsresetall

\section{Introduction}

% definition and usage of quantum network
Quantum networks enhance communication technology by transmitting and manipulating of qubits between remote locations \cite{kozlowski2019towards}. Qubits in quantum network can be sent through a wave guide such as optical fibers, or through free space \cite{van2013}. Recent research in the long-distance transmission of quantum information are almost done via optical fibers due to relatively low absorption, decoherence and fairly easily detection \cite{simon2017towards}. 
Applications of quantum networks include detecting eavesdropping in Quantum Key Distribution (\text{QKD}) \cite{bennett2014quantum}, distributed quantum computing \cite{cacciapuoti2019} and distributed quantum sensing \cite{guo2020distributed}.
% Applications of quantum networks include detecting eavesdropping in Quantum Key Distribution (\text{QKD}), distributed quantum computing and distributed quantum sensing.

% entanglement in quantum repeater network
To establish a quantum network capable of transmitting quantum information between two end nodes, a fundamental step is to create entanglement between them \cite{ursin2007entanglement}. Entanglement, which is the correlation between remote qubits, can be established between two end nodes by consuming entangled pairs. Due to the no-cloning theorem  \cite{nielsen_chuang_2010} in quantum mechanics, qubits cannot be copied. To extend the potential range of entanglement beyond the maximum distance between interconnected quantum nodes, repeaters are utilized \cite{van2008system}.
%Refs. \cite{pirandola2019end, van2008system, pant2019, van2013designing, pirandola2016capacities}. 
% In this paper, we assume to use maximally entangled pairs, known as Bell pairs, in the quantum repeater network.
% Entanglement, which is the correlation between remote qubits \cite{bernien2013heralded, moehring2007entanglement}, can be established between two end nodes by consuming entangled pairs.

% request scheduling
The field of quantum networks is still in its early stages, with even single-hop communications presenting significant challenges. 
A known feature of entanglement is that it is inherently fragile and has a probability of failure. Therefore, a challenge arises in the network layer: determining which pairs of end nodes can be serviced within the current time slot, while others are queued for subsequent time slots. This issue is known as request scheduling in quantum networking. 

Within this area, researchers are actively investigating entanglement requests scheduling methods within quantum networks. In \cite{le2022dqra}, the authors apply deep reinforcement learning methods to allocate quantum channels for accommodating multiple entanglement requests. However, the success rates of qubit entanglement and qubit swapping within the input states of the models are not considered. In \cite{cicconetti2021}, the authors integrate a quantum simulator with network simulation to investigate scheduling methods for entanglement requests. Some of the trade-offs involved between efficiency and fairness of different scheduling policies were highlighted. However, there is still a need for exploring scheduling methods that balance the trade-offs involved in scheduling decisions and further evaluation in this area, particular using more user-centric performance metrics. 

% \subsection{Contribution}
In this paper, we present a Deep Q-Network (DQN) framework applied to quantum network entanglement request scheduling that considers the request delay and fairness. The DQN framework is evaluated using a system model framework that encompasses quantum behavior and network simulation, showing its flexibility to train schedulers that manages the trade-offs between the delay and fairness.

% \subsection{Structure of the Paper}
% \textbf{[LESTER: We can remove this section to save space if needed] }The paper is structured as follows:  Section \ref{sec:sys} presents the system model. Section \ref{sec:algo} explains the scheduling methods of entanglement requests. Section \ref{sec:perf} presents the evaluation results of the proposed scheduling methods and Section \ref{sec:conc} concludes the paper.

\section{System Model} \label{sec:sys}

%The dynamic system model integrates quantum deployments with network simulations. 
To capture quantum behaviors, such as fidelity, entangled pair generation, and noise in quantum channels, a quantum simulation framework is developed. 
The NetSquid quantum simulator \cite{coopmans2021netsquid} is employed to generate distributions of fidelity and the time required to complete the entanglement of quantum links. For the network layer, a time slot simulation is used to extract features of network behavior, including the queuing, execution, and dropping of requests.

In brief, the number of entanglement requests are assigned dynamically with random source and destination nodes using a time slot simulation. Lookup tables are then employed to assign the quantum behavior involved in generating, transmitting, and measuring the entangled pairs. These lookup tables are derived from quantum simulations to obtain fidelity and the time required to complete the entanglement process between two end nodes.

\subsection{Quantum Model}
% For a single quantum state, we denote it as $|\varphi\rangle$. For multi-qubit states, we have $|\Psi\rangle = \sum_{x \in \{0,1\}^n} \alpha_x |x\rangle$, where $\sum_x |\alpha_x|^2 = 1$. 
% Some multi-qubit states can be separated into individual single-qubit states. For example, qubit $A$ in state $|0\rangle_A$ and qubit $B$ in state $|1\rangle_B$ can be written as $|01\rangle_{AB} = |0\rangle_A |1\rangle_B$. Otherwise, the multi-qubit states are entangled, meaning they are correlated with each other, such as the Bell pair $|\Phi\rangle$ shown in the Equation (\ref{eqn:bell_pair}). Once qubit $A$ is measured, the state of qubit $B$ can be determined based on the measurement result of qubit $A$.
A Bell pair, which is a pair of entangled qubits, $|\Phi\rangle$ shown in the Equation (\ref{eqn:bell_pair}) are generated in the quantum source. Once qubit $A$ is measured, the state of qubit $B$ can be determined based on the measurement result of qubit $A$.
\begin{equation}
\label{eqn:bell_pair}
|\Phi\rangle=\frac{1}{\sqrt{2}}\left(|0\rangle_A|0\rangle_B+|1\rangle_A|1\rangle_B\right)
\end{equation}
% After generating entanglement between two end nodes, we need
To quantify whether the entanglement is successful or not, fidelity \cite{van2014} $F$ is used to calculate the difference between the density matrix $\rho$ of the state that receive at the node and expected state $|\psi\rangle$.

\begin{equation}
F(|\psi\rangle, \rho) =\langle\psi|\rho| \psi\rangle
\end{equation}

% In two-node entanglement, quantum sources capable of generating Bell pairs are required. After generating the entangled pairs at one node, one of the qubits from bell pair is sent to the other end node. The fidelity of the entanglement between the two nodes is then calculated and recorded based on the measurement results obtained from both nodes.

In two-node entanglement, Bell pairs are generated by quantum sources at one end node. Therefore, one of the qubits from bell pair is sent to the other end node. The fidelity of the entanglement between the two nodes is then calculated and recorded based on the measurement results obtained from both nodes.

% Unlike two-node entanglement, repeaters and entanglement swapping are used to extend the range of quantum communication, that is limited by losses in the fiber. 
% Quantum repeaters facilitate the transmission of qubits over long distances by employing entanglement swapping. 
In long distance entanglement, distance-dependent losses in the channel becomes a limitation, and quantum repeaters are utilized to perform entanglement swapping. Entanglement swapping involves combining two short-distance Bell pairs to create a single longer-distance Bell pair \cite{van2013}. 

% \begin{figure}[H]
% \centering
% \includegraphics[width=0.85\linewidth]{figures/swapping.png}
% \caption{\label{fig:swapping} Entanglement swapping.}
% \end{figure}

\begin{figure}
\centering
\includegraphics[width=0.9\linewidth]{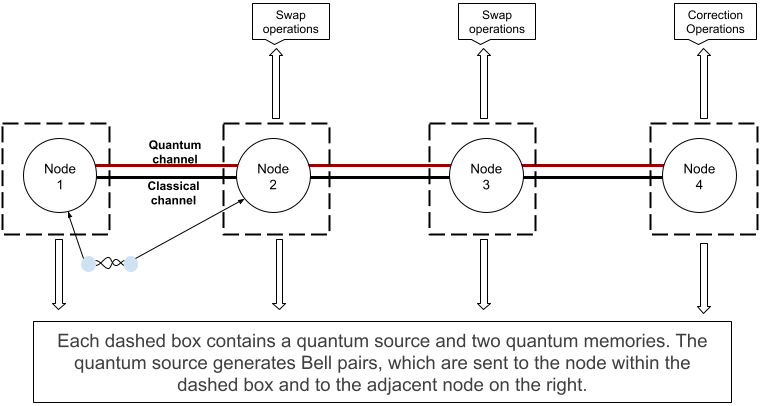}
\vspace{-0.2cm}
\caption{\label{fig:node_chain} Components of establishing entanglement in a node chain.}
\end{figure}

In the node chain, as shown in Fig. \ref{fig:node_chain}, entanglement swapping and correction processes involve three main steps: First, the intermediate node generates entanglement with its adjacent neighbors. Second, the intermediate node performs a swap operation using a projective measurement. Finally, the outcomes of this measurement are transmitted through a classical channel to the end node, which then uses an appropriate sequence of $X$ and $Z$ gates to make corrections to the local qubit at the end node.

To simplify the model, the physical deployment of repeaters is treated the same as end nodes and structure the quantum model to consist of nodes and channels. Within the nodes, physical models are implemented, including entangled pair generation, entanglement swapping, Bell state measurement, and correction. Entanglement swapping is implemented at the intermediate node and correction operations at the destination node. By comparing the obtained state with the expected state, the fidelity of this node chain entanglement is calculated.

Apart from entanglement swapping and correction, the configuration for nodes and channels are the same for both two-node and node-chain entanglement. The specific configurations are described below.

First, the quantum source that generates Bell pairs randomly samples both the correct state and the wrong state to replicate a source fidelity of $0.9$. Second, each node contains one quantum processor, which includes quantum memories and physical instructions. Third, for the error model, depolarizing and dephasing models are assigned to the quantum memory and the gate's physical instructions respectively. The parameters for the quantum deployment are given in Table \ref{tab:quatum_parameters}.
% In the quantum channel, we incorporate both a fiber delay model and a fiber loss model. Specifically, the fiber delay model represents transmission delay based on the constant speed of photons through the fiber. The fiber loss model pertains to the loss of qubits within the channel. On the other hand, in the classical channel, where we assume the transmission of classical information is perfect, we only consider the fiber delay model. The parameters for the quantum deployment is given in Table \ref{tab:quatum_parameters}.
% For the channels, we assume the classical channel is perfect, while the quantum channel follows a depolarizing error model
\begin{table}[ht]    
\vspace{-0.2cm}
\centering
\caption{\label{tab:quatum_parameters}Parameters in each node in the quantum simulation.}
\vspace{-0.2cm}
%\begin{tabular}{\columnwidth}{|l|l|l|}
\begin{tabular}{|m{3.5cm} |m{2.5cm}| }
  \hline
 Description & Value \\
 \hline
 Source fidelity & 0.9  \\
 \hline
 Number of quantum memory & 2  \\
 \hline
 Memory depolarizing rate & 6000Hz \\
 \hline
 Gate dephasing rate & 5000Hz\\
 \hline
\end{tabular}
\vspace{-0.2cm}
\end{table}

\subsection{Network Model} \label{sec:networkmodel}

In quantum networks, the interplay between end nodes exhibiting quantum behavior and time-dependent traffic at the network layer makes the analysis challenging. To incorporate the model, which includes physical models with time-dependent network behavior, a network simulator was designed, that periodically generates arrival requests and calculates execution times for building entanglements between two end nodes.

The quantum simulation is used to construct lookup tables that contains the distribution of link fidelity and execution time. For instance, to establish entanglement between designated source and destination nodes, a fidelity value is generated from these distributions. This procedure is equivalent to deploying quantum entanglement among these nodes, including the generation and distribution of Bell pairs and necessary quantum operations. If the fidelity exceeds the threshold of $0.5$, the entanglement is assumed to be successful \cite{kozlowski2020designing}, and the execution time is recorded as the final duration required to achieve entanglement. 
%If the fidelity falls below the threshold, we cumulatively add the execution times and continue to randomly select fidelities from the distribution until encountering one that surpasses the defined threshold.
If the fidelity falls below the threshold, the entanglement is assumed to be unsuccessful and another entanglement is attempted. The execution times of all the attempts are cumulatively added and until encountering an attempt that surpasses the defined threshold.
However, if the cumulative execution time exceeds a defined maximum execution time, the entanglement request is aborted, and the next request will be handled, if any exists.

The network topologies are generated randomly. However, to simplify the procedure of processing entanglement requests by referring to a lookup table, we assume that the links between the nodes in the network are homogeneous, i.e. the link distance and quality are the same.

% Second, the network simulator has the capability to generate random network topologies. However, to simplify the procedure of processing entanglement requests by referring to a lookup table, we make the following assumptions about the network topology:

% \begin{itemize}
%     \item The distance between every pair of neighboring nodes is the same. Neighboring nodes are defined as nodes that are directly connected by one hop.
%     \item The link quality between every pair of neighboring nodes is the same, meaning there is no distinction between good or bad links.
% \end{itemize}

\begin{figure}[ht]
\centering
\includegraphics[width=0.9\linewidth]{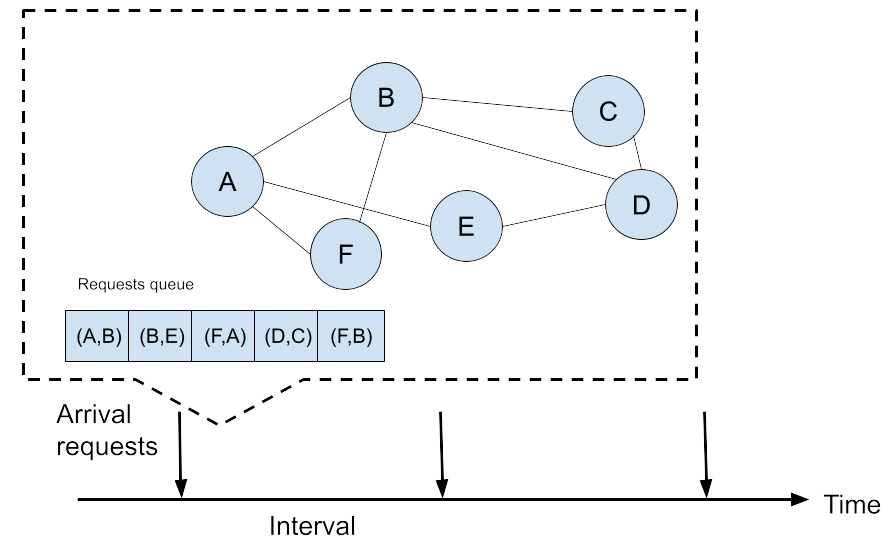}
\vspace*{-0.35cm}
\caption{An example network topology and the time slot simulation.}
\label{fig:topology}
\end{figure}

% add correct distribution for arrival request rate
To compose entanglement requests, source and destination pairs are first selected randomly, and then entanglement is established between them. The number of requests arriving at each node is also randomly generated, following a uniform distribution $U[a, b]$, where $a$ and $b$ are the lowest and highest number of requests, respectively. 
% This means that the number of requests arriving per time slot falls within the interval $[a, b]$:

% \begin{equation}
% \label{eqn:uniform_distribution}
% f(x) =
%   \begin{cases} 
%    \frac{1}{b-a} & \text{for } a \le x \le b \\
%    0       & \text{otherwise}
%   \end{cases}
% \end{equation}

% add figure for time slot explaination
Finally, as shown in Fig. \ref{fig:topology}, the network simulator incorporates the concept of time slots. At the beginning of each time slot, several new entanglement requests arrive. The duration of each time slot is fixed, and should be selected based on factors such as the network topology, quantum hardware, and the specific quantum network application. At the end of the time slot, any unfinished entanglement requests are carried over to the next time slot and queued before the newly generated requests in the subsequent time slot. Therefore, the network simulator can execute, queue, and drop requests, which could be used to evaluate the performance of different entanglement requests scheduling methods.

%  \begin{figure}[h]
% \centering
% \includegraphics[width=0.9\linewidth]{figures/time_slot.png}
% \caption{Time slots simulation}
% \label{fig:time_slot}
% \vspace{-0.2cm}
% \end{figure}

\subsection{Problem Formulation}

Given a quantum network topology $G$ with a set of entanglement requests $D$, a novel scheduling scheme for these requests, employing a Deep Q-Network (DQN) model is proposed. This scheme is designed to optimize both the delay time and fairness among the entanglement requests $D$, within the framework of network quality-of-service (QoS). The delay time is used as a measure to capture the throughput of the system, with throughput being inversely related to delay time (i.e., throughput = 1/delay time). Fairness is evaluated during each request's waiting time for constructing entanglement, given that the requests are assumed to be processed sequentially.

 \begin{enumerate}
     \item \textbf{Delay Time for Each Request:} \\ Given the assumption of a limited number of quantum memories, entanglement requests are queued for resource allocation to establish entanglement between source and destination nodes. If the request is not dropped due to exceeding the maximum execution time, the delay time for this request is calculated by subtracting the time at which the request is fulfilled from the time it was generated. The delay time $t_{d}$ for each request is calculated as
    \begin{equation}
    \label{eqn:request_delay_time}
        t_{d} = t_{f} - t_{g},
    \end{equation}
    
    where $t_{f}$ denotes the time of request fulfillment and $t_{g}$ indicates the time of request generation.
     
     \item \textbf{Jain's Fairness Index for the Generated Request Set:} \\ The Jain's fairness index, $J$, for the delay time $d$ in each request $(d_1, d_2, \ldots, d_n)$ is calculated as
     \begin{equation}
        J(d_1, d_2, \ldots, d_n) = \frac{\left( \sum_{i=1}^{n} d_i \right)^2}{n \sum_{i=1}^{n} d_i^2}.
    \end{equation}
    The range of Jain's fairness index $J$ is $(0,1)$. A higher index value indicates that delay time for each request is closer to each other, while a lower value indicates greater disparity. 
 \end{enumerate}

Within quantum networks, this paper addresses the optimization problem focusing on balancing the trade-off between minimizing delay time and maximizing fairness among entanglement requests. Therefore, scheduling methods for these entanglement requests are proposed and analyzed.

\section{Entanglement scheduling methods} \label{sec:algo}
In this section, the proposed DQN scheduling approach is described, along with the greedy, proportional fair and first in, first out (FIFO) scheduling approaches that will serve as the benchmark.
The proposed \text{DQN} scheduling method is used to balance two conflicting objectives: minimising delay time, and maximising fairness among these requests.

\subsection{\text{DQN} scheduling}
\label{sec: dqn_scheduling}
The \text{DQN} approach has discrete states and actions. The state, action and the reward are defined as follows:

\begin{enumerate}
    \item \textbf{State:} A binary matrix $D$ of size $k \times 2|v|$ is used to represent the source and destination nodes of incoming entanglement requests for the input state, where $k$ is the number of requests in the arrival request set:
    
    \begin{equation}
    \label{eqn:requests}
    D=\left[\begin{array}{cc}
    v_{s_1} & v_{d_1} \\
    v_{s_2} & v_{d_2} \\
    \vdots & \vdots \\
    v_{s_{k}} & v_{d_{k}}
    \end{array}\right].
    \end{equation}

    $v$ is a binary vector $v=\left[\left\{v_i ; i=1, \ldots, |V|\right\}\right]$ representing the positions of the source $v_{s}$ or destination $v_{d}$ nodes in a request, as 

    \begin{equation}
    \label{eqn:single_request}
    v_i= \begin{cases}1, & \text{if the current node is } i \\ 0, & \text{otherwise}\end{cases}.
    \end{equation}
    
    Therefore, each row of the matrix (\ref{eqn:requests}) represents a request, where the positions of $1$ correspond to the node numbers in the graph for the source or destination nodes. If a request is resolved, its source and destination representing row would be replaced with $0$-vectors. 
    
    Moreover, if the number of arrival requests $k$ is fixed per time slot, the size of the request set matrix $D$ is fixed since the length of rows depends on the network topology, which remains unchanged during training.  
    
    \item \textbf{Action:} A discrete action space is utilized in which the model is trained to select the pending request with the highest score at each step until all requests are processed. At each step $t$, the DQN scheduler selects the action 
    
    \begin{equation}
    \label{eqn:action}
    a^{(t)} = \operatorname{argmax}_i \left( r_i^{(t)} \right),
    \end{equation}
    
    corresponding to the highest reward $r^{(t)}$ among all pending requests. The reward set, which contains the rewards for each pending request, is defined as $r^{(t)}=\left\{r_0^{(t)}, r_1^{(t)}, \ldots, r_{|D|}^{(t)}\right\}$.
\vspace{0.1cm}
    \item \textbf{Reward:} The reward is assigned to balance the delay time, and fairness among these requests. The specific reward is designed as follows:
    \begin{equation}
    \label{eqn:delay_time}
        r_{1}= (\text{min}_{d} - \text{cur}_{d}) / \text{max}_{d}
    \end{equation}
    \begin{equation}
    \label{eqn:jain_index}
        r_{2} = c_{j} - 1
    \end{equation} 
    \begin{equation}
    \label{eqn:reward}
        r = c_{d} \times r_{1} + c_{j} \times r_{2} 
    \end{equation}

    $\text{min}_{d}$, $\text{cur}_{d}$, and $\text{max}_{d}$ represent the minimum, current, and maximum total delay time among these requests respectively, where $\text{min}_{d}$ and $\text{max}_{d}$ are calculated using the execution time of each request after all the requests are executed.
    
    The parameters $c_{d}$ and $c_{j}$ represent the coefficients of the reward from delay time and Jain's index, respectively. These two parameters are used to adjust the proportion of the reward attributed to minimizing delay time versus fairness. 
    % For example, setting $c_{d}$ to $0.9$ and $c_{j}$ to $0.1$ means that the \text{DQN} will prioritize minimizing delay time, resulting in queuing the request with the least execution time for building an entanglement first.
    For example, setting $c_{d}$ to $0.9$ and $c_{j}$ to $0.1$ means that the \text{DQN} will prioritize minimizing delay time. Conversely, setting $c_{d}$ lower than $c_{j}$ means that the \text{DQN} will focus more on maximizing fairness among these requests.   
\end{enumerate}

\begin{figure}[ht]
\centering
\includegraphics[width=0.9\linewidth]{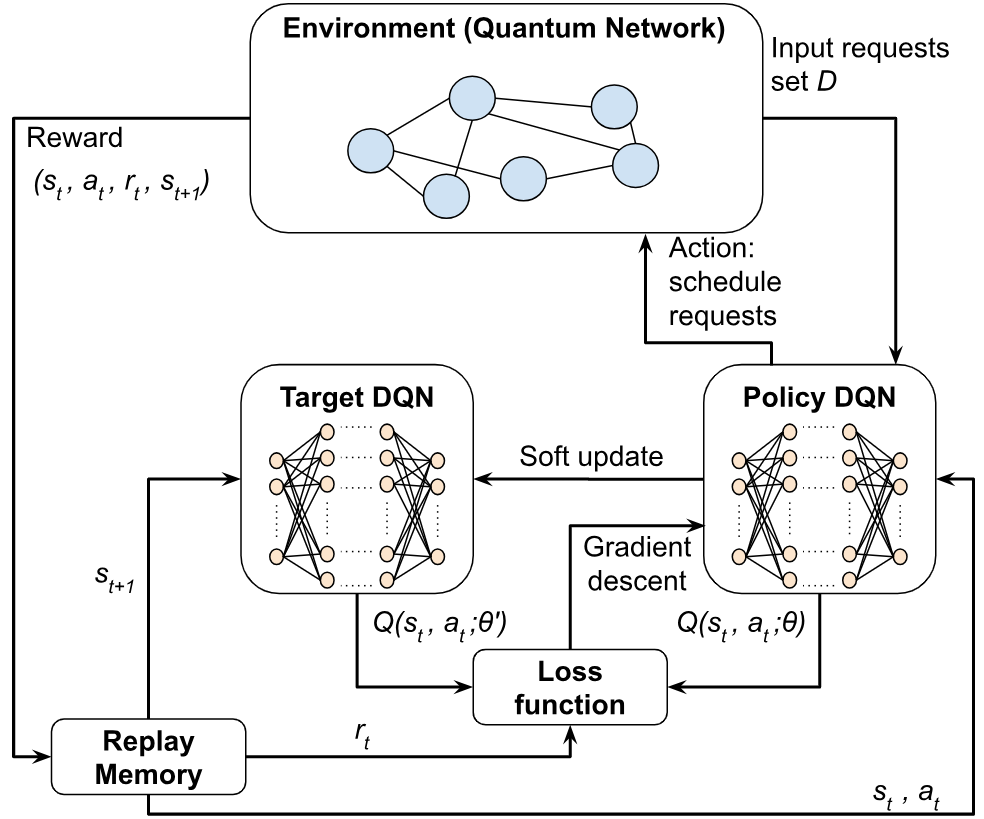}
\vspace{-0.05cm}
\caption{Architecture of the Double DQN method in simulations.}
\label{fig:dqn_structure}
\end{figure}

To stabilize the learning process of DQN, Double DQN is considered, as shown in Fig. \ref{fig:dqn_structure}. The policy DQN is the main DQN model used to interact with the environment, determining which request to select at each time step. The target DQN is employed for stabilization. Both the policy and target DQN calculate the $Q$-value $Q$ based on current state $s_{t}$, action $a_{t}$ and parameters $\theta$ or $\theta'$ of their neural network, which enables the model to make decisions by selecting actions that maximize the expected future rewards. 

\begin{figure}[ht]
\centering
\includegraphics[width=0.9\linewidth]{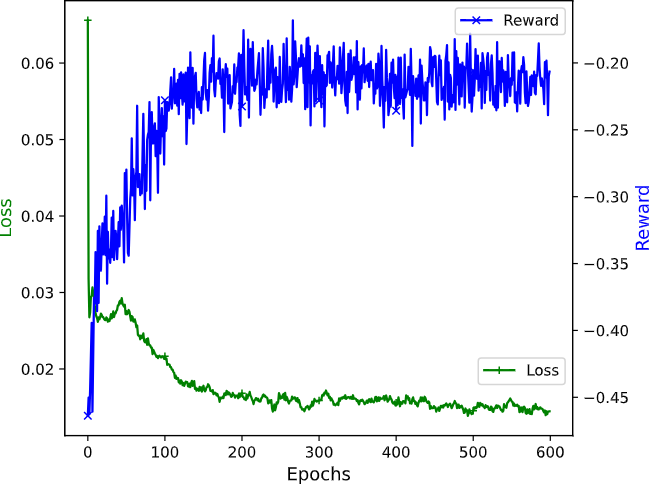}
\vspace{-0.25cm}
\caption{Loss and reward values in $5$ requests Double DQN model.}
\label{fig:dqn_loss_reward}
\vspace{-0.7cm}
\end{figure}

In Fig. \ref{fig:dqn_loss_reward}, we trained the Double DQN model to output request sets with an arrival of $5$ requests per time slot, biased toward minimizing delay time with $c_{d}$ to $0.9$ and $c_{j}$ to $0.1$. Each epoch represents the average of $150$ samples.

%\subsection{Greedy and proportional fair scheduling}

\subsection{Greedy scheduling}
In Greedy scheduling, the source and destination pairs that have the smallest distances are selected first. Since the time required to establish entanglement between two nodes is typically shorter for smaller distances, this approach has the advantage of reducing the delay time for each request.

{\centering
\begin{minipage}{.9\linewidth}
\begin{algorithm}[H]
\caption{Greedy Scheduling}
\KwIn{SourceDestinationPairs}
\KwResult{Scheduled requests with minimized delay}
Sort requests by distance in ascending order\;
\\
\For{each request in requests}{
    \begin{itemize}
        \item Select the request with the smallest distance\;
    \end{itemize}
}
    \begin{itemize}
        \item Output ordered requests for processing\;
    \end{itemize}
\end{algorithm}
\end{minipage}
\par
}

\subsection{Proportional fair scheduling}

Proportional fair scheduling aims to achieve fairness among requests by giving higher priority to those potentially need longer time to establish one entanglement, which correlates with paths have longer distance. The mechanism for proportional fair ensures that when a request is randomly selected, those with potentially longer time to execute have a greater likelihood of being chosen. Once a request is completed, the remaining requests initiate a new round of the selection procedure until all requests are selected. Therefore, proportional fair scheduling does not guarantee absolute fairness but strives to be fair most of the time.

{
\centering
\begin{minipage}{.9\linewidth}
\begin{algorithm}[H]
\caption{Proportional Fair Scheduling}
\KwIn{SourceDestinationPairs}
\KwResult{Fairly scheduled requests}
\While{remaining requests}{    
    \begin{itemize}
        \item Calculate requests' likelihood of being chosen\;
        \item Select a request on likelihood\;
        \item Remove the selected request\;
    \end{itemize}   
}
\begin{itemize}
    \item Output ordered requests for processing\;
\end{itemize} 
\end{algorithm}
\end{minipage}
\par
}

\subsection{FIFO scheduling}

The FIFO scheduling serves the requests in the order they appear in the generated request set. It has an element of randomness as the source-destination pairs of the requests are generated randomly, and the scheduling does not prioritize fairness or delay. FIFO scheduling is used here as a benchmark to show the effectiveness of the above proposed three scheduling methods. 

%The FIFO scheduling does not prioritize fairness or delay. It serves the requests in the order they appear in the generated request set which is random since the arrival requests are generated randomly ane these requests are arrival simultaneously. FIFO scheduling is used to show the effectiveness of the above proposed three scheduling methods. 
% Therefore, the FIFO scheduling is based on the default generated request set and the output has the feature of randomness. 

\section{Performance evaluation} \label{sec:perf}
 
In this section, the numerical results for the performance of the scheduling methods are presented. The evaluation is based on the delay time for each request and Jain's Fairness Index for the request sets.

\subsection{Simulation Settings}
A random Watts-Strogatz graph $G(V,K)$ \cite{watts1998collective} where $V = 10$, $K = 3$, and $p = 0.6$ is used for the simulations. This means the generation of the graph starts with a ring lattice with $10$ vertices and $3$ edges per vertex, followed by the rewiring of each edge with a probability of $p = 0.6$. Source and destination pairs of arrival requests are randomly selected. At the start of each time slot, a certain number of requests arrive. The network simulation runs over a total of $10,000$ time slots. As described in Section \ref{sec:networkmodel}, if a request is not completed within the current time slot, it is queued for the next slot, but prioritized ahead of newly arriving requests. The maximum execution time allowed for establishing an entanglement between end nodes is $100,000\,ns$. If this time is exceeded, the request is aborted. 

As described in Section \ref{sec: dqn_scheduling}, the size of the input state matrix $D$ is fixed, requiring multiple DQNs to be trained to handle different numbers of requests. Consequently, three distinct DQN models are trained to handle $3$, $4$, and $5$ arrival requests. If the number of arrival requests falls within $[0,2]$, the model behaves the same as the Greedy or Proportional fair method, depending on the DQN method's bias. The simulation parameters mentioned above are summarized in Table \ref{tab:par}.
%As described in Section \ref{sec: dqn_scheduling}, the size of the input state matrix $D$ is fixed, meaning that the number of arrival requests per time slot must remain constant. Consequently, three distinct DQN models are trained to handle scenarios with $3$, $4$, and $5$ arrival requests. If the number of arrival requests falls within $[0,2]$, the model behaves similarly to the Greedy or Proportional Fair method, depending on the DQN method's bias. The simulation parameters mentioned above are summarized in Table \ref{tab:par}.

\begin{table}[ht]
%\vspace{0.25cm}
\centering
\caption{Network simulations parameters.}
\vspace{-0.2cm}
%\begin{tabular}{\columnwidth}{|l|l|l|}
\begin{tabular}{| m{1.2cm}  |m{3.5cm} |m{2.5cm}| }
  \hline
 Parameter & Description & Value \\
 \hline
 $V$ & Number of nodes & 10  \\
 \hline
 $G$ & Watts-Strogetz graph & K = 3, p = 0.6  \\
 \hline
 $N_t$ & Number of time slots & 10,000 \\
 \hline
 $E_m$  & Max execution time & 100,000\,ns\\
 \hline
 $D_a$  & Number of arrival requests & $U[0,5]$\,per time slot\\
 \hline  
\end{tabular}
\label{tab:par}
\end{table}

\subsection{Performance Analysis}

\textbf{Time slot interval analysis for scheduling methods}: 

The arrival request model follows a uniform distribution $U[0, 5]$, resulting in an average of $2.5$ arrival requests per time slot. 

% For the entanglement process, the maximum execution time is $10^{5}\,ns$, and the average execution time per request is approximately $5 \times 10^{4}\,ns$. Thus, we define the extreme congested case as $1.5 \times 10^{5}\,ns$, the congested case as $2 \times 10^{5}\,ns$, and the non-congested case as $5 \times 10^{5}\,ns$. This classification ensures that even if all requests reach their maximum execution time, they can still be completed within a single time slot.

% \begin{figure}[ht]
% \centering
% \includesvg[width=0.9\linewidth]{figures/1.5e5_mod.svg}
% \caption{Delay time for completing entanglement requests with time interval $1.5 \times 10^{5}\,ns$ (high load traffic).}
% \label{fig:1.5e5}
% \end{figure}
\begin{figure}[ht]
\centering
\includegraphics[width=0.9\linewidth]{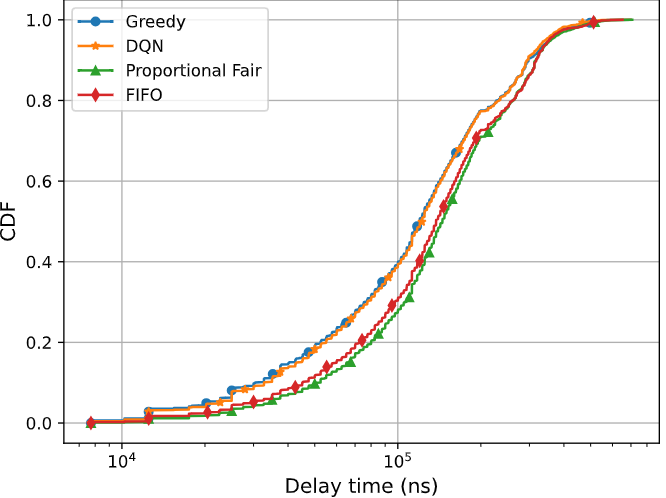}
\vspace{-0.25cm}
\caption{Delay time for completing entanglement requests with time slot interval $2 \times 10^{5}\,ns$ (medium load traffic). DQN biased towards minimizing delay.}
\label{fig:2e5}
\end{figure}
\begin{figure}[ht]
\centering
\includegraphics[width=0.9\linewidth]{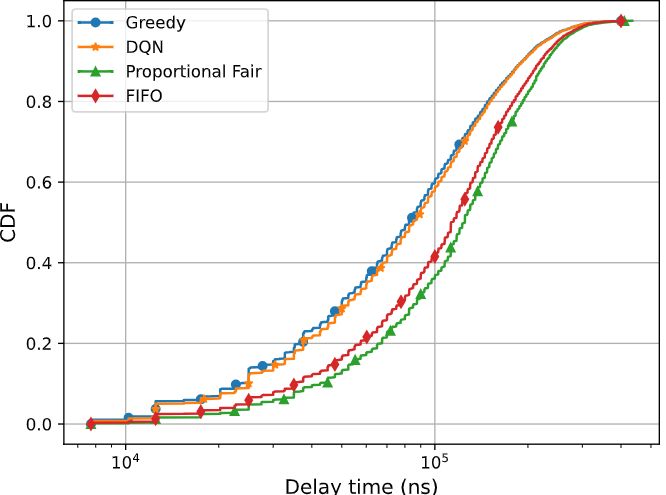}
\vspace{-0.25cm}
\caption{Delay time for completing entanglement requests with time slot interval $5 \times 10^{5}\,ns$ (low load traffic). DQN biased towards minimizing delay.}
\label{fig:5e5}
\vspace{-0.4cm}
\end{figure}
% In Fig. \ref{fig:1.5e5}, the network simulation is under high-load traffic conditions, resulting in congested conditions where most of the requests arriving per time slot cannot be executed within the same time slot. Consequently, there is little difference in the delay times between the three scheduling methods. 

In high-load traffic conditions, the network nodes are congested where most of the requests arriving per time slot cannot be executed within the same time slot. Consequently, there is little to no difference in the delay times between the different scheduling methods.

Figures \ref{fig:2e5} and \ref{fig:5e5} shows the cumulative distribution function (CDF) of the entanglement request completion delay for medium and low traffic loads, respectively. In medium and low loads, most requests are completed within their designated time slots with fewer pending cases. This allows the scheduling methods to show varying behaviors.%In the low load case, the maximum execution time is limited to $10^{5}\,ns$, allowing all generated requests to be completed within their time slots. This results in smoother curves compared to the other two figures.
%Fig. \ref{fig:2e5} shows the cumulative distribution function (CDF) of the entanglement request completion delay for medium-load traffic, where most requests are completed within their designated time slots with fewer pending cases. This allows the scheduling methods to show varying behaviors. In Fig. \ref{fig:5e5}, the maximum execution time is limited to $10^{5}\,ns$, allowing all generated requests to be completed within their time slots. This results in smoother curves compared to the other two figures.

The results for the proposed DQN approach, shown in Fig. \ref{fig:2e5} and Fig. \ref{fig:5e5} uses reward coefficients $c_d$ and $c_j$ set to $0.9$ and $0.1$, respectively, giving a bias towards minimizing delay over fairness. In this scenario, the DQN approach achieves request delay times that are similar to Greedy method. 
However, the fairness of the DQN approach surpasses that of the Greedy method. 

Given that Jain's fairness is unitless, we use the difference between the lowest and highest fairness achieved to normalize the gain calculation. The normalized gain that the DQN approach achieves over Greedy is at least 14\%, i.e. ($J_{DQN}-J_{Greedy})/(J_{max}-J_{min})$.

% This is because the greedy strategy consistently places the request with the maximum execution time at the end, leading to a large delay time for the final request. Although proportional fairness prioritizes the maximum execution time request first, it is affected by the most severe congestion issues compared to the other two approaches.

% 2e5
% jains indexs are: greedy 0.4704247599847083; dqn 0.4892584395789858; proportional 0.5775895517079145; fifo 0.5457075914838846
% percentiles in greedy are: 25 17942.38742498934; 50 39999.23808068782; 75 95440.85666850043
% percentiles in dqn are: 25 20221.316408983752; 50 45000.734365954995; 75 100220.46423599057
% percentiles in proportional are: 25 34998.82168721071; 50 75002.0060347137; 75 135001.97367191545
% percentiles in fifo are: 25 25337.69126409525; 50 65003.147343297416; 75 125001.12725836807
% Gain in fairness 0.2501723064771225

% 5e5
% jains indexs are: greedy 0.5173107116883867; dqn 0.5323396131077169; proportional 0.6182549177828603; fifo 0.5870217633275521
% percentiles in greedy are: 25 12501.098402185366; 50 32721.380766549788; 75 67499.86417686846
% percentiles in dqn are: 25 17499.18021672964; 50 35220.09658686665; 75 70500.21528079134
% percentiles in proportional are: 25 27501.114664964378; 50 65000.96719041467; 75 117943.24547591203
% percentiles in fifo are: 25 24998.692051718186; 50 52502.31013190746; 75 107497.14039818197
% Gain in fairness 0.21558850520749598
\vspace{-0.4cm}
\begin{table}[ht]
    \centering
    \caption{Jain's indexes with DQN biased towards minimizing delay.}
    \vspace{-0.2cm}
    \begin{tabular}{|c|c|c|c|c|}
        \hline
        Time slot interval & Greedy & DQN & FIFO & P. fair\\
        \hline
        $2 \times 10^{5}\,ns$ (medium load)& 0.4704 & 0.4893 & 0.5457 & 0.5776\\
        \hline
        $5 \times 10^{5}\,ns$ (low load)& 0.5173 & 0.5323 & 0.5870 & 0.6183\\
        \hline
    \end{tabular}
    \label{tab:jains_indexs_table_1}
\end{table}

% \begin{figure}[ht]
% \centering
%  \includesvg[width=0.8\linewidth]{figures/fairness_bar_chart_mod.svg}
% \caption{Percentage of DQN outperforms greedy in Jain's fairness index.}
% \label{fig:fairness_bar_chart}
% \end{figure}

\begin{figure}[ht]
\centering
\includegraphics[width=0.9\linewidth]{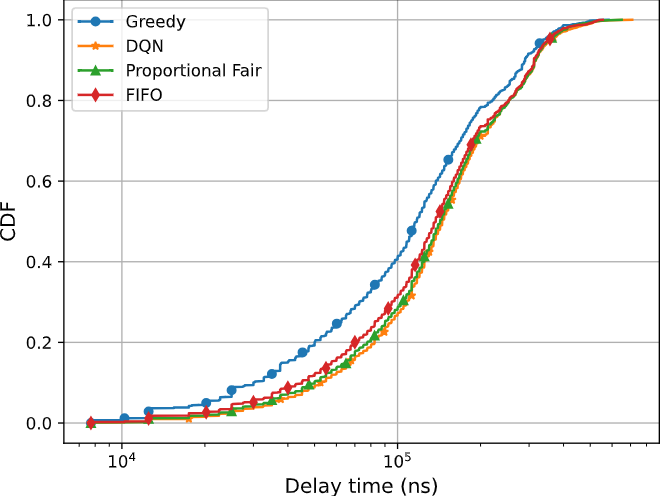}
%\caption{The DQN reward as comprising $15\%$ from reducing the delay time and $85\%$ for fairness with time slot interval $2 \times 10^{5}\,ns$.}
\vspace{-0.25cm}
\caption{Delay time for completing entanglement requests with medium load traffic, with DQN biased towards maximizing fairness.}
\label{fig:2e5_0.85f}
\vspace{-0.25cm}
\end{figure}

\textbf{Flexibility of tuning between delay time and fairness}: 

% In Fig.\ref{fig:2e5_0.85f} shows the results when $c_d$ is $0.15$, and $c_j$ is $0.85$. In the case of higher fairness, the $\text{DQN}$ achieves a higher Jain's index compared to the proportional fairness method. Finding the fairest combination of requests is a factorial problem in mathematics. Specifically, if there are $3$ requests to be ranked, there will be $6$ possible combinations. Consequently, all these different fairness values need to be calculated for finding the most fair combination. For more requests, the number of possible combinations increases significantly. The $\text{DQN}$ provides a way to achieve fairness in scheduling requests within the network after training.
To demonstrate the flexibility of the DQN approach, the rewards are changed to have a bias towards fairness, with $c_d$ and $c_j$ set to $0.15$ and $0.85$ respectively.
%In Fig. \ref{fig:2e5_0.85f}, the $\text{DQN}$ approach achieves fairness when $c_d$ is $0.15$ and $c_j$ is $0.85$. 
% In this case, the $\text{DQN}$ achieves a higher Jain's fairness index compared to Proportional fair, as shown in Table \ref{tab:jains_indexs_table_2}, but similar performance in delay time as Proportional fair scheme, as shown in Fig. \ref{fig:2e5_0.85f}. Since the time slot interval for both experiments in Fig. \ref{fig:2e5} and Fig. \ref{fig:2e5_0.85f} is $2 \times 10^{5}\,ns$, these figures illustrate the flexibility of $\text{DQN}$ in balancing delay time and fairness.
In this case, the normalized gain in fairness of the DQN approach over the Proportional fair method, ($J_{DQN}-J_{PFair})/(J_{max}-J_{min})$, is 12.8\%, as shown in Table \ref{tab:jains_indexs_table_2}. However, DQN has similar performance in delay time as Proportional fair scheme, as shown in Fig. \ref{fig:2e5_0.85f}. These results illustrate the flexibility of $\text{DQN}$ in balancing delay time and fairness.

% In table \ref{tab:percentiles}, The similarity between DQN/Proportional fair and FIFO/Proportional fair remains higher in the $25th$ to $75th$ percentiles. Since the time slot interval for both experiments in Fig. \ref{fig:2e5} and Fig. \ref{fig:2e5_0.85f} is $2 \times 10^{5}\,ns$, these figures illustrate the flexibility of $\text{DQN}$ in balancing delay time and fairness.

% Furthermore, $\text{DQN}$ achieves a $24\%$ improvement in Jain's fairness index compared to proportional fairness, as shown in Table \ref{tab:jains_indexs_table_2}.

% 0.9
% greedy 0.5013146782511927; dqn 0.6262454926840327; proportional 0.6033387686900011
% \begin{table}[ht]
%     \centering
%     \begin{tabular}{|c|c|c|c|}
%         \hline
%         time slot interval & greedy & dqn & proportional\\
%         \hline
%         $2 \times 10^{5}\,ns$ & 0.50131 & 0.62624 & 0.60333\\
%         \hline
%     \end{tabular}
%     \caption{Table of jains indexs with higher reward from fairness}
%     \label{tab:jains_indexs_table}
% \end{table}

% 0.85
% jains indexs are: greedy 0.4762087134685801; dqn 0.592687764861195; proportional 0.577789259451887; fifo 0.5458588708516681
% percentiles in greedy are: 25 17942.38742498934; 50 38163.703947433765; 75 92724.05474318241
% percentiles in dqn are: 25 35001.90775784105; 50 80004.05613463698; 75 137720.2530900528
% percentiles in proportional are: 25 32942.26877905434; 50 72503.0722790584; 75 134997.91197966784
% percentiles in fifo are: 25 25002.919690100476; 50 62939.97346970583; 75 124998.84208272942
% 0.85
% jains indexs are: greedy 0.47702842431443476; dqn 0.5994671554860428; proportional 0.5757878972983053
\vspace{-0.5cm}
\begin{table}[ht]
    \centering
    \caption{Jain's indexes with DQN biased towards maxmizing fairness.}
    \vspace{-0.2cm}
    \begin{tabular}{|c|c|c|c|c|}
        \hline
        Time slot interval & Greedy & DQN & FIFO & Proportional fair\\
        \hline
        $2 \times 10^{5}$$\,ns$ & 0.4762 & 0.5927 & 0.5459 & 0.5778\\
        \hline
    \end{tabular}
    \label{tab:jains_indexs_table_2}
\end{table}

% \begin{table}[ht]
%     \centering
%     \caption{Table of percentiles with DQN biased towards fairness.}
%     \begin{tabular}{|c|c|c|c|c|}
%         \hline
%         Percentiles & Greedy & DQN & FIFO & Proportional fair\\
%         \hline
%         $25th$ & 17942$\,ns$ & 35001$\,ns$ & 25002$\,ns$ & 32942$\,ns$\\
%         \hline
%         $50th$ & 38163$\,ns$ & 80004$\,ns$ & 62939$\,ns$ & 72503$\,ns$\\
%         \hline
%         $75th$ & 92724$\,ns$ & 137720$\,ns$ & 124998$\,ns$ & 134997$\,ns$\\
%         \hline
%     \end{tabular}
%     \label{tab:percentiles}
% \end{table}
\vspace{-0.3cm}
\section{Conclusion and Future Work} \label{sec:conc}
 
In this paper, a \text{DQN}-based scheduler for entanglement requests in a quantum network is proposed and evaluated using network simulations with periodically generated requests, and compared with benchmark methods. The benefit of the proposed \text{DQN} approach is its ability in managing the trade-offs between delay time and fairness. It is shown that when configured with a bias towards minimizing delay, the proposed DQN approach was able to train schedulers that achieved the same level of low delays, but with higher fairness compared with Greedy schedulers. Conversely, the DQN was also able to achieve higher fairness compared to Proportional fair schedulers when trained with a bias towards fairness. During the training process, the reward mechanism learns from the execution time of completing an entanglement request, which makes it more flexible compared to other scheduling methods.

% In future work, the joint optimization of scheduling with resource allocation with variable link reliability will be considered. In this more complex problem, DQN and other AI-based approaches will lead to achieve even better performance.

In future work, the joint optimization of maximizing entanglement rates with the constraints of limited entangled pairs, heterogeneous links and quantum memory multiplexing within the nodes will be considered. Reinforcement learning techniques, such as Double DQN and Proximal Policy Optimization, will also be explored, potentially offering improved stability and performance.
\vspace{-0.2cm}
\section*{Acknowledgements}
This publication has emanated from research conducted with the financial support of Research Ireland under Grant number \text{13/RC/2077\_P2}. For the purpose of Open Access, the author has applied a CC BY public copyright licence to any Author Accepted Manuscript version arising from this submission.
%This work was funded by Science Foundation Ireland under the CONNECT Phase 2 project, grant number \text{13/RC/2077\_P2}.
\vspace{-0.4cm}
\bibliographystyle{IEEEtran}
\bibliography{ref}
\end{document}